# Double-sided van der Waals epitaxy of topological insulators across an atomically thin membrane

Joon Young Park[1,2], Young Jae Shin[1], Jeacheol Shin[2], Jehyun Kim[2], Janghyun Jo[3], Hyobin Yoo[1,4], Danial Haei[1,5], Chohee Hyun[6], Jiyoung Yun[2], Robert M. Huber[7,8], Arijit Gupta[7,8], Kenji Watanabe[9], Takashi Taniguchi[10], Wan Kyu Park[8], Hyeon Suk Shin[11,12], Miyoung Kim[3], Dohun Kim[2], Gyu-Chul Yi[2,*], and Philip Kim[1,*]

[1] Department of Physics, Harvard University, Cambridge, MA 02138, USA
[2] Department of Physics and Astronomy, and Institute of Applied Physics, Seoul National University, Seoul 08826, Republic of Korea
[3] Department of Materials Science and Engineering, and Research Institute of Advanced Materials, Seoul National University, Seoul 08826, Republic of Korea
[4] Department of Physics, Sogang University, Seoul 04107, Republic of Korea
[5] Center for Nanoscale Systems, Harvard University, Cambridge, MA 02138, USA
[6] Department of Chemistry, and Low Dimensional Carbon Materials Center, Ulsan National Institute of Science and Technology (UNIST), Ulsan 44919, Republic of Korea
[7] Department of Physics, Florida State University, Florida, 32306, USA
[8] National High Magnetic Field Laboratory, Florida State University, FL 32310, USA
[9] Research Center for Electronic and Optical Materials, National Institute for Materials Science, 1-1 Namiki, Tsukuba 305-0044, Japan
[10] Research Center for Materials Nanoarchitectonics, National Institute for Materials Science, 1-1 Namiki, Tsukuba 305-0044, Japan
[11] Department of Energy Science and Department of Chemistry, Sungkyunkwan University (SKKU), Suwon 16419, Republic of Korea
[12] Center for 2D Quantum Heterostructures, Institute of Basic Science (IBS), Sungkyunkwan University (SKKU), Suwon 16419, Republic of Korea
[*] e-mail: gcyi@snu.ac.kr; pkim@physics.harvard.edu

**Atomically thin van der Waals (vdW) films provide a novel material platform for epitaxial growth of quantum heterostructures. However, unlike the remote epitaxial growth of three-dimensional bulk crystals, the growth of two-dimensional (2D) material heterostructures across atomic layers has been limited due to the weak vdW interaction. Here, we report the double-sided epitaxy of vdW layered materials through atomic membranes. We grow vdW topological insulators (TIs) $Sb_2Te_3$ and $Bi_2Se_3$ by molecular beam epitaxy on both surfaces of atomically thin graphene or hBN, which serve as suspended 2D vdW "*substrate*" layers. Both homo- and hetero- double-sided vdW TI tunnel junctions are fabricated, with the atomically thin hBN acting as a crystal-momentum-conserving tunnelling barrier with abrupt and epitaxial interface. By performing field-angle dependent magneto-tunnelling spectroscopy on these devices, we reveal the energy-momentum-spin resonant tunnelling of massless Dirac electrons between helical Landau levels developed in the topological surface states at the interface.**

Two-dimensional (2D) van der Waals (vdW) heterostructures composed of materials with various electronic properties serve as novel platforms to explore a variety of interesting physical properties and device applications[1,2]. One of the most intriguing systems is a vertical heterostructure consisting of two materials separated by an atomically thin insulating barrier, allowing vertical tunnelling transport through vdW interfaces[3-5]. Such a heterostructure with high crystal quality and epitaxial alignment is of particular interest because crystal momentum parallel to the interface can be conserved, offering a unique platform for investigating the intrinsic selection rules of the resonant tunnelling process[6-8]. The surface states of three-dimensional (3D) vdW topological insulators (TIs) of $(Bi, Sb)_2(Te, Se)_3$ compounds[9], when integrated into vdW tunnel junctions, can provide exciting opportunities to discover novel quantum phenomena occurring at the atomically sharp epitaxial interface between topologically dissimilar atomic layers[10-15].

Multilayered TI/normal insulator vdW heterostructures have been constructed by sequential direct growth using molecular beam epitaxy (MBE)[10,16]. In this conventional MBE growth, the choice of the barrier materials is constrained by chemical compatibility and lattice constant matching. Furthermore, the common occurrence of step edges during the growth process significantly reduces the effectiveness of forming a smooth vdW heterointerface.

VdW epitaxy on both surfaces of an atomically thin, single-crystalline suspended vdW membrane opens a new way to construct TI epitaxial heterostructures with atomically sharp interfaces. This novel synthesis route presented in this work distinguished from the recently demonstrated remote epitaxy of 3D/2D/3D hybrid systems. In remote epitaxial growth, the electrostatic field of the substrate (lower 3D layer) penetrates the 2D vdW interlayer (Fig. 1a)[17,18]. Due to its strong dependence on the ionic character or polarity of the 3D substrate, remote epitaxy establishes the long-range epitaxial relationship between 3D/3D bulks, insensitive to the crystallographic alignment at the 3D/2D local interfaces. On the contrary, in the newly proposed double-sided vdW epitaxy, as shown in Fig. 1b, one can utilize the short-range local vdW interaction at the interfaces, structurally connecting the top and bottom epilayers to the substrate (middle) layer. Moreover, unlike remote epitaxy where a stronger coupling leads to a stronger correlation of the electronic structure across the heterostructure, the electrical coupling can be tuned independently by the thickness and electronic properties of the middle layer. To date, while 3D/2D/3D quasi-vdW epitaxial double heterostructures have been demonstrated in the form of graphene sandwiched by semiconductor nanostructures with reconstructed dangling bonds of polar facets at the interface[19,20], all-2D double-sided vdW epitaxy has not yet been realized. In this Article, we present the double-sided vdW-MBE growth of 2D/2D/2D structures where two epitaxially aligned TIs are separated by atomically thin crystalline barriers with abrupt interfaces, which has enabled us to observe the resonant tunnelling between Landau levels (LLs) developed in the topological surface states (TSSs) at the vdW heterointerface where the energy, momentum, and spin of the helical Dirac fermions are conserved during the tunnelling process.

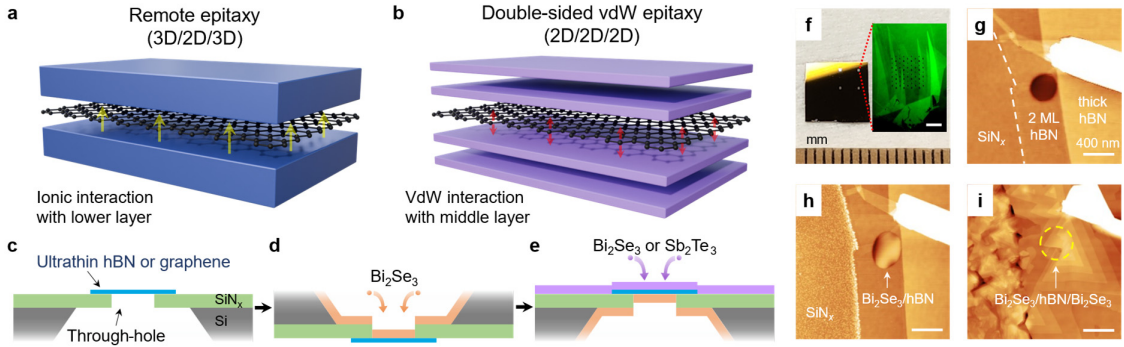

**Fig. 1. Double-sided vdW epitaxy of TI/hBN(graphene)/TI vertical heterostructures. a,b**, Schematics illustrating the growth mechanism of remote epitaxy (**a**) and double-sided vdW epitaxy (**b**). In remote epitaxy, the middle 2D vdW layer serves as an atomically thin spacer to mediate the long-range interaction between 3D crystals. In the new concept of double-sided vdW epitaxy, the middle vdW layer serves as an atomically thin growth template where the top and bottom layers are epitaxially connected by the short-range vdW interaction. **c–e**, Schematics of the double-sided vdW epitaxy procedure, where TIs are grown on the bottom and top surfaces of atomically thin vdW substrates suspended on perforated $SiN_x$ membranes. **f**, Photograph of an array of $SiN_x$ membrane windows compatible with MBE, TEM, and microfabrication. Inset: High-contrast optical microscope image of an array of holes fabricated in the $SiN_x$ membrane covered by an atomically thin hBN crystal. Scale bar: 10 μm. **g–h**, AFM topographic images (20 nm colour scale) of 2 ML hBN suspended over one of the holes in the $SiN_x$ membrane taken before (**g**) and after (**h**) the first growth of $Bi_2Se_3$ on the back side, and after the second growth of $Bi_2Se_3$ on the front side (**i**).

The method developed to fabricate the double-sided epitaxial vdW heterostructures is schematically illustrated in Fig. 1c–e. Ultrathin hBN layers (2–5 monolayers; MLs) or monolayer graphene are suspended over hole arrays pre-patterned in $SiN_x$ membranes to expose both surfaces (Fig. 1f and Extended Data Fig.1). Subsequently, $Bi_2Se_3$ (n-type) and $Sb_2Te_3$ (p-type) thin films are heteroepitaxially grown on each side of the suspended layer using ultra-high vacuum (UHV) MBE (see Methods). The suspended vdW layer serves as a crystalline substrate for the epitaxy of the TI films. In particular, the atomically thin hBN layers can serve as a wide bandgap insulating barrier for tunnelling spectroscopy between the two TIs whose interface is sealed under UHV. Since the suspended vdW heterostructures are electron-transmitting and mechanically robust, this templated growth approach enables us to directly use plan-view TEM and atomic force microscopy (AFM) to study their structural properties (Fig. 1g–i and Fig. 2) and the microfabrication processes to form electrical contacts to the junctions[21-24].

The microstructural properties of the TI/hBN(graphene)/TI heterostructures can be investigated by TEM. We first demonstrate that bottom (b-) and top (t-) $Bi_2Se_3$ films are epitaxially aligned through a suspended monolayer graphene, a single-atom-thick vdW crystal substrate. Figure 2a shows the plan-view selected area electron diffraction (SAED) image of the resulting heterostructure exhibiting the preferential epitaxial relationship of $\{10\bar{1}0\}_{b\text{-}Bi_2Se_3} \parallel \{10\bar{1}0\}_{graphene} \parallel \{10\bar{1}0\}_{t\text{-}Bi_2Se_3}$. The weak Bragg peaks from $Bi_2Se_3$ with small deviation angles can be attributed to the wrinkles formed in the suspended vdW layers (inset of Fig. 2a and ref.[23]) and to grains with a small in-plane misorientation angle, which is likely due to the weak vdW interactions at the interface[22,25-27].

The same methodology can be applied to insulating hBN substrates, which is of particular interest for vertical tunnel junctions. The cross-sectional HR-TEM image in Fig. 2b shows the quintuple layer (QL) structures of the $Bi_2Se_3$ thin films, and the atomically clean and abrupt interfaces between b-$Bi_2Se_3$, 2 ML hBN, and t-$Bi_2Se_3$ (see also Extended Data Fig. 2). Their epitaxial relationship can be confirmed by comparing the fast Fourier transform (FFT) of the representative areas (yellow and orange boxes) with the interfacial region (red box), where one can find matching structures of b- and t-$Bi_2Se_3$ (white arrows) relative to the hBN atomic registry (red arrows).

Next, we present the growth of dissimilar TI materials—namely $Bi_2Se_3$ and $Sb_2Te_3$—across a suspended hBN, further demonstrating the power of our double-sided vdW epitaxy. Figure 2c shows the plan-view HR-TEM image of the $Sb_2Te_3$/hBN/$Bi_2Se_3$ heterostructure, taken near the edge of a hole in the $SiN_x$ membrane so that there is a region where only $Bi_2Se_3$ is grown on hBN (first growth on the top side) without the coverage of $Sb_2Te_3$ (second growth on the bottom side). While the moiré patterns arise from the lattice mismatch between $Sb_2Te_3$ and $Bi_2Se_3$ (~2.8%) (refs.[28,29]) and their small in-plane misorientations (Fig. 2d), the FFT patterns of the selected areas of the image (Fig. 2d,e) identify the epitaxial alignment between $Bi_2Se_3$ and hBN as well as the epitaxial alignment between the $Bi_2Se_3$ and $Sb_2Te_3$ films through the hBN substrate: $\{10\bar{1}0\}_{Bi_2Se_3} \parallel \{10\bar{1}0\}_{hBN} \parallel \{10\bar{1}0\}_{Sb_2Te_3}$. The epitaxial relationship in our bidirectional MBE on suspended hBN and graphene is consistent with that of the TI thin films grown on a single surface of hBN and graphene substrates (Extended Data Fig. 3 and ref.[22,27]), suggesting that the epitaxial alignment is mainly due to the interaction with the suspended vdW substrate layer.

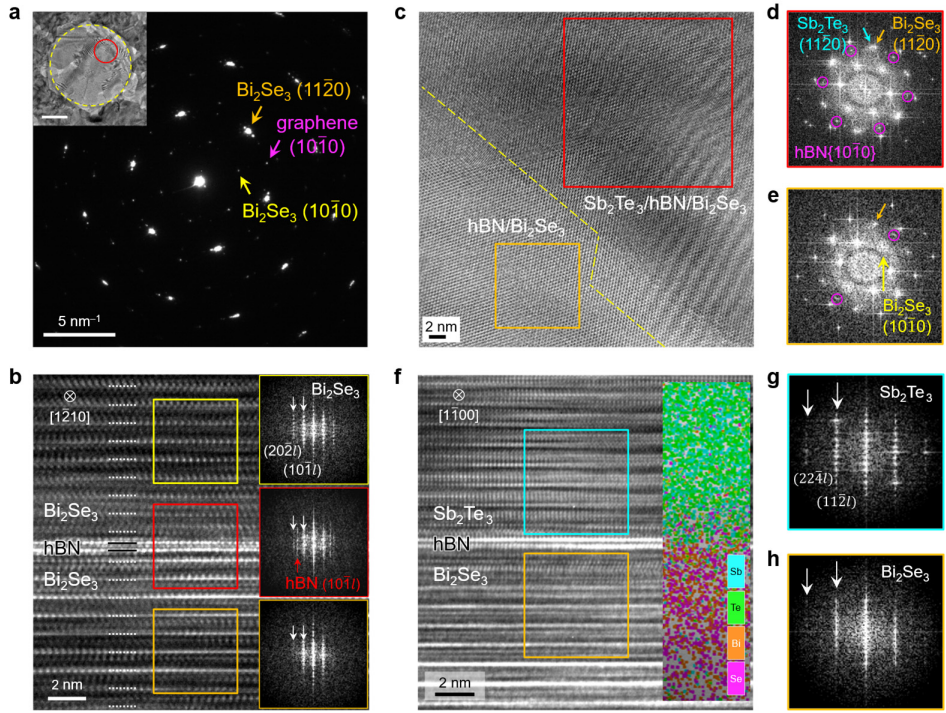

**Fig. 2. Structural properties of TI/hBN(graphene)/TI double-sided vdW epitaxial heterostructures. a**, Plan-view SAED pattern of $Bi_2Se_3$/monolayer graphene/$Bi_2Se_3$. Inset: Plan-view bright-field (BF) TEM image of the heterostructure. The dashed yellow circle and the red circle indicate the suspended area (a hole in the $SiN_x$ membrane) and the area in which the SAED is recorded, respectively. Scale bar: 200 nm. **b**, Cross-sectional HR-TEM image of $Bi_2Se_3$/2ML hBN/$Bi_2Se_3$ taken along the zone axis $[1\bar{2}10]$. Insets: FFT patterns of the areas corresponding to the yellow, red, and orange boxes in the main panel. **c**, Plan-view HR-TEM image of $Sb_2Te_3$/few-layer hBN/$Bi_2Se_3$ taken around the boundary of $Sb_2Te_3$ (dashed yellow line). **d,e** FFTs of the regions marked by the red (**d**) and orange (**e**) boxes in **c**. **f**, Cross-sectional HR-TEM image of $Bi_2Se_3$/2ML hBN/$Sb_2Te_3$ taken along the zone axis $[1\bar{1}00]$. Inset: Cross-sectional STEM image overlaid with EDS map. **g,h** FFTs of the regions marked by cyan (**g**) and orange (**h**) boxes in **f**.

The high interfacial quality of the heterostructure is also confirmed by the cross-sectional HR-TEM (Fig. 2f). Since the strong and dense in-plane chemical bonding in the hBN layers provides a completely impermeable atomic membrane, we do not expect any interdiffusion of atoms through the hBN substrate during the growth of the second layers on the opposite side of the surface after the first layer growth. The inset of Fig. 2f shows cross-sectional chemical composition analysis using scanning tunnelling electron microscopy (STEM) equipped with energy-dispersive X-ray spectroscopy (EDS). The signals of Sb and Te are observed only above the hBN layer while those of Bi and Se are detected only below the hBN layer, demonstrating the clear spatial separation of $Sb_2Te_3$ and $Bi_2Se_3$ across the two-atom-thick hBN barrier. The epitaxial relationship across the hBN substrate can be again confirmed by comparing the FFT patterns of $Bi_2Se_3$ and $Sb_2Te_3$. As shown in Fig. 2g,h, the $(11\bar{2}l)$ and $(22\bar{4}l)$ spots of $Bi_2Se_3$ (white arrows in Fig. 2g) align well with the corresponding spots of Sb$_2$Te$_3$ (white arrows in Fig. 2h).

With the hBN barrier layer, our double-sided vdW epitaxy provides TI/hBN/TI tunnelling devices, enabling the study of electron tunnelling properties of helical Dirac fermions through the tunnelling barrier. To perform tunnelling spectroscopy, individual electrical contacts to an array of suspended junctions are fabricated on the top side while they share a common bottom contact (Fig. 3a; see also Extended Data Fig. 4 and Methods). A DC bias voltage $V$ is applied to the top electrode with respect to the bottom electrode, and the tunnelling current $I$ can be measured across the junction. Although we have fabricated tunnelling devices on Bi$_2$Se$_3$/hBN/Bi$_2$Se$_3$ as well (results in Extended Data Fig. 5), in the remainder of this paper, we focus on the Bi$_2$Se$_3$ (bottom)/hBN/Sb$_2$Te$_3$ (top) heterostructure, whose configuration yields n-TI/insulator/p-TI junctions.

The Bi$_2$Se$_3$/hBN/Sb$_2$Te$_3$ heterojunction at zero bias is expected to form a type-III band alignment with a broken gap due to the large difference in their electron affinities and relatively small bulk band gaps (Fig. 3b)[30,31]. Accordingly, charge transfer and band bending result in carrier accumulation near the junction. Away from the heterointerface, the Fermi level $E_F$ of Bi$_2$Se$_3$ (Sb$_2$Te$_3$) films is located near the edge of the bulk conduction (valence) band due to the unintentional n-type (p-type) doping[22,31,32]. The TSSs at the interface are hence electrically contacted by the bulk states.

Figure 3c shows $J(V)$, the current density $J$ as a function of the applied bias voltage $V$, measured in devices with different hBN barriers ranging from 2 to 5 ML. The measured differential conductance at zero bias, normalized by the junction area $G_0 = \partial I/\partial V(V=0)$, scales exponentially with the thickness of the insulating hBN barrier (inset of Fig. 3c), suggesting that electron transport in our devices is dominated by tunnelling through hBN barriers[33,34]. For a negative bias voltage, $|J(V)|$ monotonically increases rapidly with increasing $|V|$. However, for the positive bias side, $J(V)$ increases non-monotonically as $V$ increases, exhibiting negative differential conductance-like characteristics at a certain bias voltage range. These features are consistent with the characteristics of tunnel diodes formed by p–n junctions with broken bandgap alignments[35].

The detailed tunnelling process in the junction can be described in the differential conductance $G(V) = \partial I/\partial V$ curve for the device with a 2 ML hBN barrier (Fig. 3d). Schematic band alignments at the heterointerface under different bias voltages are illustrated together, as labeled by (i) – (iv). The differential conductance exhibits a strongly reduced value between bias voltage regions (i) and (iv), indicating that the Fermi level of n-Bi$_2$Se$_3$ ($E_F^{BS}$) crosses the bulk band gap of p-Sb$_2$Te$_3$ as $V$ is increased in this region. As the Fermi level of p-Sb$_2$Te$_3$ ($E_F^{ST}$) moves out of the BCB of n-Bi$_2$Se$_3$ and into its bulk band gap ((ii) and (iii)), the $G(V)$ exhibits a peak at $V = 225$ mV. This local maximum can be attributed to the enhanced tunnelling density of states (DOS) when the bulk band edges align in the presence of band bending[36-38]. At $V \lesssim 40$ mV and $V \gtrsim 350$ mV, oscillatory spectral features appear due to the bulk quantum well states (QWSs)[38-40].

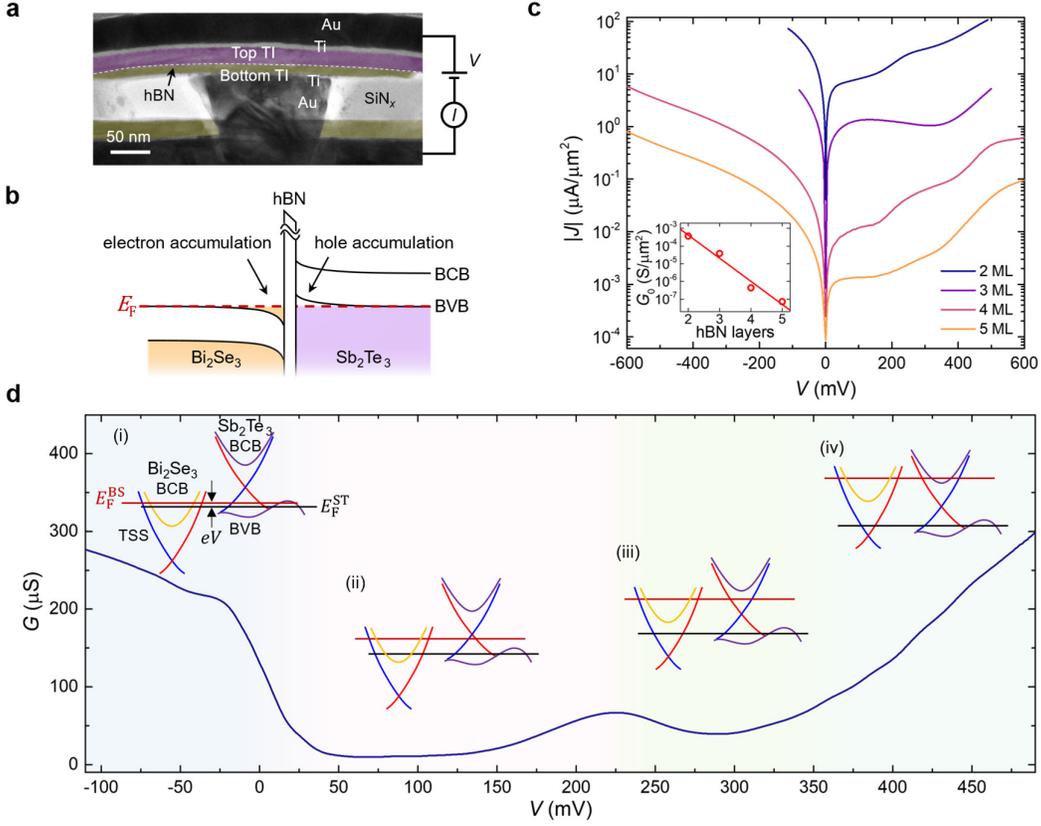

**Fig. 3. Tunnelling spectroscopy in Bi$_2$Se$_3$/hBN/Sb$_2$Te$_3$ junctions at zero magnetic field. a**, False-coloured low-magnification cross-sectional BF TEM image of a vertical tunnel junction device in two-terminal measurement configuration. While this image shows Bi$_2$Se$_3$/hBN/Bi$_2$Se$_3$, the device structure is identical for all measured junctions, including Bi$_2$Se$_3$/hBN/Sb$_2$Te$_3$, except for the hBN barrier thickness and the junction area. **b**, Schematic band alignment of the Bi$_2$Se$_3$/hBN/Sb$_2$Te$_3$ junction at zero bias voltage, depicted in energy – out-of-plane axis plane. **c**, Tunnelling current density as a function of bias voltage for the devices with different thicknesses of hBN barrier. $V$ is applied to Sb$_2$Te$_3$ (top), and the current drained through Bi$_2$Se$_3$ (bottom), $I$, is measured. Inset: Zero-bias differential conductance per unit area as a function of the number of hBN layers. **d**, Differential conductance of the Bi$_2$Se$_3$/2ML hBN/Sb$_2$Te$_3$ device as a function of $V$. (i) – (iv): Schematic band alignments at the tunnelling interface under different bias conditions, plotted in the energy – in-plane momentum plane. TSSs are marked by red and blue colours, indicating their spin-momentum locking textures. The tunnelling conductance is determined by the position of the bulk conduction band (BCB) and the bulk valence band (BVB) relative to the $E_F^{BS}$ and $E_F^{ST}$. Devices with 2 and 3 ML hBN barriers are measured at temperature $T = 1.8$ K and those with 4 ML and 5 ML barriers are measured at $T = 1.5$ K.

The topological nature of the coupled surface states in our vdW epitaxial structure can be revealed by the magneto-tunneling transport study. When magnetic field perpendicular to the junction, $B_\perp$, is applied, LLs form in the TSSs, which can modulate the tunnel conductance associated with the TSSs[40-43]. Figure 4a shows the differential conductance spectra $G(V, B_\perp) = \partial I/\partial V(V, B_\perp)$ of the Sb$_2$Te$_3$/2ML hBN/Bi$_2$Se$_3$ device at different $B_\perp$. As $B_\perp$ increases, the tunnelling conductance spectrum exhibits additional oscillatory features. A more systematic analysis can be performed by subtracting off the zero magnetic field conductance: $\Delta G(V, B_\perp) = \partial I/\partial V(V, B_\perp) - \partial I/\partial V(V, B_\perp = 0)$. Figure 4b shows $\Delta G$ as a function of $V$ and $B_\perp$. We observe that $\Delta G(V)$ oscillations exhibit larger amplitudes at higher $B_\perp$. These oscillatory features can be categorized into three groups: oscillations with peak and valley positions (I) that are independent of $B_\perp$; (II) that vary with $B_\perp$ and converges to $V \sim 135$ mV as $B_\perp$ vanishes; and (III) that vary with $B_\perp$ and converges to a large bias voltage ($V > 450$ mV) as we extrapolate the position of the origin.

To further investigate the nature of $\Delta G$ oscillation, we tilt the magnetic field, $B$, and provide a component of the magnetic field parallel to the junction, $B_\parallel$. Figure 4b–e shows $\Delta G(V, B)$ at different $B_\perp$ and $B_\parallel$. We find that the oscillatory feature (I) depends only on the total magnetic field $B$, indicating a 3D isotropic nature, while features (II) and (III) depend only on $B_\perp$, suggesting a 2D origin. We attribute the feature (I) to the bulk QWSs due to its angle-insensitivity and the bias voltage ranges in which the peaks appear (see Extended Data Fig. 6).

The 2D oscillatory features (II) and (III) can be related to the resonant tunnelling associated with LL formation in the TSSs. In particular, the observed unequal spacing, which increases with increasing $B_\perp$, suggests that these LLs originate from the linearly dispersed Dirac spectrum of the TSSs. In a perpendicular magnetic field, the Dirac dispersion of the TSSs can lead to sharply quantized LLs at energy $E_n = E_D + \text{sgn}(n)v_F\sqrt{2e\hbar|n|B_\perp}$, where $E_D$ is the energy of Dirac point (DP), $v_F$ is the Fermi velocity of the TSSs, $e$ is the elementary charge, $\hbar$ is the reduced Planck constant, and $n = 0, \pm1, \pm2, ...$ is the LL index, ignoring the Zeeman energy[40-43]. As the bias voltage increases, $V > 20$ mV, the resonant tunnelling between the BCB of Bi$_2$Se$_3$ and the LLs developed in the TSS of Sb$_2$Te$_3$ becomes possible, and indeed, the oscillatory features (II) in the magneto-tunnelling spectra Fig. 4b–d follow $e(V - V_0) = \text{sgn}(n)v_F\sqrt{2e\hbar|n|B_\perp}$, centered around $V_0 \approx 135$ mV. We identify the peaks of these oscillations, $V_n$, and assign the LL indices $n$ as shown in the upper part of Fig. 4b, marked by grey fonts (see Extended Data Fig. 7). We then plot the positions of the peaks in energy $eV_n$ against $\text{sgn}(n)\sqrt{|n|B_\perp}$ as shown in the lower part of Fig. 4f. Remarkably, all data points for $n > 0$ (electron LLs) and $n < 0$ (hole LLs) separately form a straight line and converge at $B_\perp = 0$ when extrapolated, within 1 mV which is our measurement resolution. The slope of each linear fit line provides the estimation of the $v_F$ of the corresponding TSS: we obtain $v_F \approx 4.1 \times 10^5$ m/s for $n < 0$ LLs (red line) and $5.6 \times 10^5$ m/s for $n > 0$ LLs (blue line), and the corresponding bands are assigned to the valence and conduction band of TSSs in Sb$_2$Te$_3$, respectively.

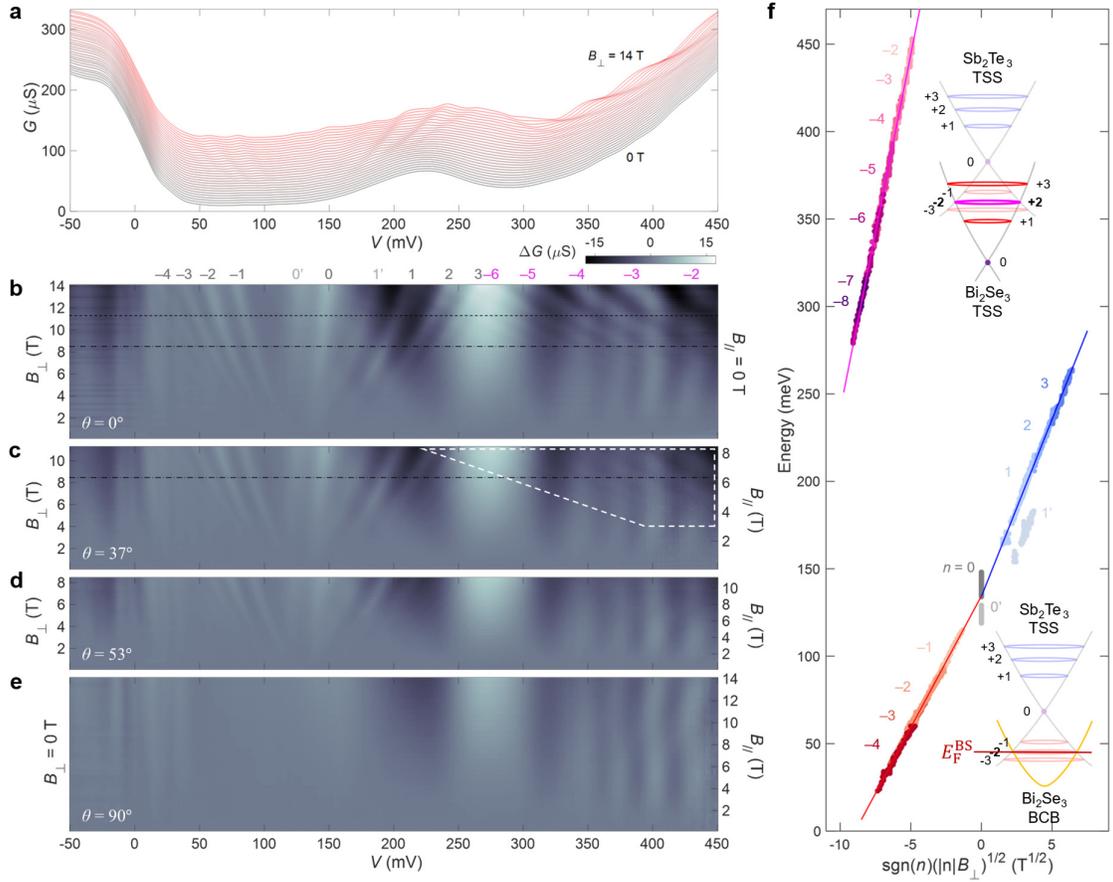

**Fig. 4. Magnetic field-dependent tunnelling conductance of the Bi$_2$Se$_3$/2ML hBN/Sb$_2$Te$_3$ device. a**, $B_\perp$-dependence of tunnelling conductance. Spectra under different fields, in 0.5 T increments, are vertically offset by 4 µS for clarity. **b–e**, Colour scale maps showing the field-modulated component of the conductance $\Delta G(V, B)$ at different field angles $\theta$ defined with respect to the out-of-plane direction. Perpendicular and in-plane components of the field are indicated by the left and right ticks, respectively. Vertical axes of **b–d** are adjusted to have the same $B_\perp$ scale, while those of **b,e** have the same $|B|$ scale. The black dashed (dash-dotted) horizontal line indicates the maximum $B_\perp$ in **c (d)**. **f**, LL peak positions at different $B_\perp$ plotted versus $\text{sgn}(n)(|n|B_\perp)^{1/2}$. The lines are linear fits to the data. The lower (upper) inset is the schematic representation of the band alignment that causes the LL peaks in the tunnelling conductance to appear at lower (higher) bias voltages. The spin helicity of the TSSs is expressed in red and blue colours. Numbers denote LL indices. All measurements are performed at $T = 1.8$ K.

There are, however, two exceptional data sets that do not fall into the LL schemes discussed above. First, the peak of the oscillation corresponding to $n = 0$, started at $V_0 \approx 135$ mV at zero magnetic field, exhibits a weak linear dispersion as $B_\perp$ increases, with an estimated $g$-factor of ~29. Applying an external magnetic field to the TSS breaks the time-reversal symmetry to open a gap, causing a Zeeman energy shift, most significantly affecting the zeroth LL[44]. The experimentally measured $g$-factors, however, vary widely among the experiments[43-45]. Second, there exists another copy of $n = 0$ and 1 LL peaks, marked as 0' and 1' respectively, emerging at lower bias voltages with smaller intensities. Although their origin is unclear, these modulations do not split from or merge with the main LL peaks.

We now turn our attention to the high bias oscillations (III) above. We identify their peak positions $V_n^*$ and assign the corresponding LL indices (Extended Data Fig. 4), marked by magenta fonts in the upper part of Fig. 4b. We then perform a similar analysis as for (II) and obtain the corresponding LL origin $V_0^* \approx 647$ mV and velocity $v \approx 1.1 \times 10^6$ m/s (magenta line in Fig. 4f). At this higher bias voltage regime, both of the $E_F^{BS}$ and $E_F^{ST}$ are outside of the BVB and BCB of the counterpart, respectively, and the resonant tunnelling can occur between the LLs developed in the TSSs of $Bi_2Se_3$ and $Sb_2Te_3$ (see the upper inset schematic in Fig. 4f). In order for such tunnelling to happen, the quantized energy, $E_n$, and angular momentum of LLs, $k_n = \sqrt{2e|n|B_\perp/\hbar}$, as well as the spin-momentum locking helicity of TSSs should be matched. Considering the two TSSs at the tunnelling interface have the opposite spin-momentum locking helicity due to their opposite surface normal directions, the only allowed inter-LL resonant tunnelling is the inter-band tunnelling between an occupied – $n^{th}$ LL of $Bi_2Se_3$ to an unoccupied $n^{th}$ LL of $Si_2Te_3$ satisfying $eV_n^* + E_D^{BS} - \text{sgn}(n)v_F^{BS}\sqrt{2e\hbar|n|B_\perp} = E_D^{ST} + \text{sgn}(n)v_F^{ST}\sqrt{2e\hbar|n|B_\perp}$, where $v_F^{BS(ST)}$ and $E_D^{BS(ST)}$ are the Fermi velocity and the DP energy of the $Bi_2Se_3$ ($Sb_2Te_3$) TSS measured with respect to the $E_F^{BS(ST)}$, respectively. This relationship results in a straight line in the plot of $eV_n^*$ versus $\text{sgn}(n)\sqrt{|n|B_\perp}$ as shown in the upper part of Fig. 4f, where the slope of this line corresponds to $v_F^{BS} + v_F^{ST}$, which is approximately twice as high as $v_F$ for TSSs $Bi_2Se_3$ and of $Sb_2Te_3$[32,40,41,43,46], in agreement with our observation. Thus, the energy-momentum-spin resonance between two sets of LLs manifests as the sum of the $v_F$'s of $Bi_2Se_3$ and $Sb_2Te_3$.

Finally, we discuss the effect of the in-plane magnetic field $B_\parallel$ on the resonant tunnelling between LLs. As shown in Fig. 4c,d, the $\Delta G$ peaks corresponding to inter-LL tunnelling (III) are rapidly suppressed with increasing $B_\parallel$ (e.g., compare the LL peaks in the region highlighted by the white dashed box in Fig. 4c with the corresponding $B_\perp$ and $V$ ranges in Fig. 4b,d). The application of $B_\parallel$ provides an in-plane momentum gain to the tunnelling electrons[6,8], which suppresses inter-band inter-LL resonant tunnelling, where the energy, angular momentum, and spin of the electrons must be conserved. Note that the tunnelling from bulk band to LL (II) is not suppressed by the presence of finite $B_\parallel$ (Fig. 4b–d), since the electron momentum is not quantized in the BCB of $Bi_2Se_3$ from which electrons tunnel.

The realization of two different types of TI materials epitaxially coupled via an ultrathin crystalline barrier, with atomically sharp and clean interfaces and without chemical intermixing, has allowed us to systematically investigate the selection rules of electron tunnelling between the TSSs. Our results show that the two TSSs can be brought together to a sub-nm distance, which is much shorter than the thickness limit of a single 3D TI slab thanks to the wide bandgap of hBN[39,40,47,48]. This system can serve as a platform to explore interesting physics involving coupled TSSs, such as the proposed topological exciton condensation[14,15]. The double-sided vdW epitaxial growth technique can be further applied to various combinations of vdW materials including superconductors and ferromagnets, potentially leading to emergent phenomena and novel device applications.

## Methods

**Preparation of suspended ultrathin hBN and graphene**

For vertical tunnel junctions, atomically thin hBN single crystal flakes are prepared by mechanical exfoliation on the oxygen-plasma treated surfaces of Si substrates with a 90 nm $SiO_2$ layer and identified by optical microscopy. The thickness and surface cleanliness of the identified layers are further characterized by AFM. We choose 2–5 ML hBN for the tunnel barrier because they provide reliably working tunnel junctions over a suitable current and bias voltage range. The target flake is then picked up and transferred onto perforated $SiN_x$ membranes using sticky polymer films, such as polycaprolactone (PCL)[49] or polycarbonate (PC), on polydimethylsiloxane (PDMS) stamps. Here, we employ 50 nm thick low-stress silicon-rich nitride membranes with TEM-compatible silicon frames (Silson Ltd). Arrays of holes with typical diameters of 0.4 – 1 μm are fabricated in the $SiN_x$ membranes by standard electron (*e*)-beam lithography and $CF_4$/Ar reactive ion etching. The oxygen plasma treatment of the $SiN_x$ membranes immediately prior to the drop-down process enhances the adhesion between the vdW layers and the membranes, reducing the risk of wash-away of the flake or wrinkles formation during the subsequent wet process. After the transfer, the sticky polymer is removed by organic solvents: tetrahydrofuran (THF) for PCL and chloroform for PC. Finally, the sample is annealed (typically at 350–400°C) under high vacuum to reduce the polymer residues. For plan-view TEM study of the double-sided vdW epitaxial heterostructures, we also employ few-layer hBN epitaxially grown on a sapphire substrate[50] and monolayer graphene grown on a Cu foil. These CVD-grown large-scale vdW substrate layers are suspended on the holey $SiN_x$ membranes by the poly(methyl methacrylate) (PMMA)-assisted wet transfer method. The suspended vdW layer undergoes another round of *in-situ* thermal cleaning in the UHV-MBE chamber before the growth of TI thin films.

**Double-sided vdW-MBE growth**

First, unless otherwise stated, $Bi_2Se_3$ thin films are typically grown first on the bottom surface of the suspended hBN or graphene through the holes in the $SiN_x$ membranes by loading the chip upside-down on a substrate holder of the MBE. We employ the two-step growth process developed for high-quality $Bi_2Se_3$ thin film growth on vdW-layered substrates[22,26]. The top surface of the suspended vdW layer remains intact after bottom side growth, preserving its atomically clean surface (see Fig. 1h). This growth result is attributed to the highly directional molecular fluxes of the precursors which are hardly subject to scattering events during MBE growth in a UHV environment. Next, the substrate is inverted *ex-situ* and loaded back to the MBE chamber so that the top side faces the molecular fluxes. Another round of two-step MBE growth is then performed to grow either $Bi_2Se_3$ or $Sb_2Te_3$ thin films on top of the hBN(graphene)/$Bi_2Se_3$ heterostructure using a similar two-step growth process[27]. In this step, care is taken to protect the pre-grown $Bi_2Se_3$ such as omitting the thermal cleaning process and reducing the 2$^{nd}$ step growth temperature. The b-TI film has crystalline structures up to the interface with hBN and exhibits similar structural quality to the t-TI film (Fig. 2b,f), indicating that the hBN layer provides excellent protection of the bottom layer from the environment during the *ex-situ* sample flipping process and subsequent growth of the top layer. The typical thickness of the TI films is 20 QL for the top side and 10 QL for the bottom side.

**Microstructural characterizations**

The TEM lamella for the cross-sectional study is prepared by a focused ion beam system (FEI Helios 650). Field-emission TEMs with an acceleration voltage of 200 kV (JEOL JEM-2100F and FEI Tecnai F20) are used for HR imaging. An average background subtraction filter is applied to remove noise in the HR-TEM images. The elemental distribution mapping is performed with a silicon drift detector-based EDS system operating in the STEM mode and analyzed with AZtec software (Oxford Instruments).

**Device fabrications**

Electrical contacts on both sides of the suspended TI/hBN/TI heterostructures, prepared on $SiN_x$ membrane templates, are fabricated by the following steps. First, Ti/Au 5/200 nm metal layer is deposited on the bottom side of the sample by *e*-beam evaporation. The thick metal layer provides ohmic contacts to the bottom TI and mechanical support to the membrane for the following fabrication steps. We note that the frames of our membrane chips are made of high-resistivity silicon, which becomes insulating at low temperatures. The individual top ohmic contacts to the array of junctions are fabricated by *e*-beam lithography followed by *e*-beam evaporation of Ti/Au 5/55 nm. Using the metal contacts as etch masks, an Ar plasma etching is carried out to remove the TI films on the uncovered area, electrically separating the array of junctions. Finally, bonding pads and interconnection lines to the contacts are deposited by another round of *e*-beam lithography and *e*-beam evaporation of Ti/Pd/Au 5/20/125 nm. The typical junction area is in the range of 0.2 to 0.7 μm².

**Magneto-tunnelling measurements**

The current–voltage characteristics of the tunnelling devices are measured using a voltage source (Yokogawa GS200 DC voltage source or Keithley Instruments 2400 SourceMeter) to apply DC bias voltage and a low-noise current preamplifier (DL Instruments Model 1211) connected to a digital multimeter (Agilent 34401A) for measuring the current. The differential conductance is measured simultaneously with the current–voltage curve by mixing the DC bias voltage with a small AC excitation voltage of 2 mV at 47.77 Hz using a lock-in amplifier (SR830, Stanford Research Systems), a voltage divider, and a transformer. Temperature and magnetic field are controlled by either Oxford Instruments Teslatron PT (base temperature $T_{base}$ = 1.5 K, maximum magnetic field $B_{max}$ = 8 T) or Quantum Design Physical Property Measurement System ($T_{base}$ = 1.8 K, $B_{max}$ = 14 T). The angle of magnetic fields with respect to the tunnelling interface is controlled by changing the orientation of sample *ex-situ*. For each field angle, tunnelling characteristics at zero magnetic field are cross-checked to confirm the robustness of the tunnelling conductance against thermal cycling. The $Bi_2Se_3/hBN/Bi_2Se_3$ device presented in Extended Data Fig. 5 is measured at the National High Magnetic Field Laboratory ($T_{base}$ = 0.3 K, $B_{max}$ = 31.5 T) with a rotator probe that can change the orientation of sample *in-situ*. During the device fabrication, loading, and shipping processes, meticulous measures are implemented to prevent electrostatic discharge.

## Data availability

The data that support the findings of this study are presented in the Article and its Extended Data. Further data are available from the corresponding authors upon reasonable request.

## Acknowledgements

We thank J. Eom, Y. S. Kim, and C.-H. Lee for their help in the transfer of atomically thin vdW crystals on the perforated $SiN_x$ membranes. The major part of this work was supported by the Office of Naval Research (ONR) Multidisciplinary University Research Initiatives (MURI) program (N00014-21-1-2377) and the Science Research Center (SRC) for Novel Epitaxial Quantum Architectures (NRF-2021R1A5A1032996). J.S., J. K., and D.K. acknowledge support from the National Research Foundation of Korea (NRF) grants funded by the Korean Government (MSIT) (RS-2023-00283291, RS-2023-00207732, and No. 2023R1A2C2005809). J.J. and M.K. acknowledge support from the NRF grant NRF-2022R1A2C3007807. H.S.S acknowledges support from the Institute for Basic Science (IBS-R036-D1). A portion of the user collaboration grant program (UCGP) project was performed at the National High Magnetic Field Laboratory supported by the National Science Foundation through NSF/DMR-1644779 and the State of Florida. K.W. and T.T. acknowledge support from the JSPS KAKENHI (Grant Numbers 20H00354 and 23H02052) and World Premier International Research Center Initiative (WPI), MEXT, Japan.

## Author contributions

P.K., G.-C.Y., and J.Y.P. conceived the experiments. J.Y.P. performed the double-sided vdW epitaxy and fabricated the tunnelling devices. J.Y.P. and G.-C.Y. analysed and optimized these processes. J.Y.P. performed the magneto-tunnelling measurements together with J.S. and J.K.. J.Y.P., R.M.H., A.G., and W.K.P. carried out the measurement at the National High Magnetic Field Laboratory. J.Y.P., J.S., J.K., D.K., and P.K. analysed the transport data. J.J. conducted the TEM experiments. J.J., J.Y.P., G.-C.Y., and M.K. analysed the TEM data. J.Y.P. and H.Y. prepared the hole-patterned $SiN_x$ membrane templates. Y.J.S., D.H.N., H.Y., and J.Y.P. carried out the suspension of ultrathin vdW layers on the perforated $SiN_x$ membranes. C. H. and H.S.S. provided the epitaxial few-layer hBN films and collaborated on discussions about the double-sided vdW epitaxy on them. J.Y. grew the monolayer graphene. K.W. and T.T. provided the bulk hBN single crystals. G.-C.Y. and P.K. jointly supervised the project. J.Y.P. and P.K. wrote the manuscript with input from all authors.

## Competing interests

The authors declare no competing interests.

# Extended data figures

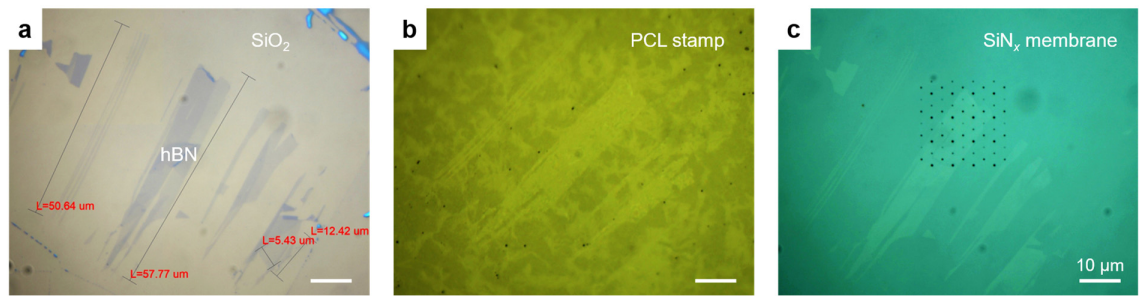

**Extended Data Fig. 1. Suspension of atomically thin hBN crystals on the perforated SiN$_x$ membrane. a–c,** Optical microscope images of ultrathin hBN layers: as exfoliated on the SiO$_2$ surface (**a**), picked up by the PCL polymer stamp (**b**), transferred on the SiN$_x$ membrane with premade holes (**c**).

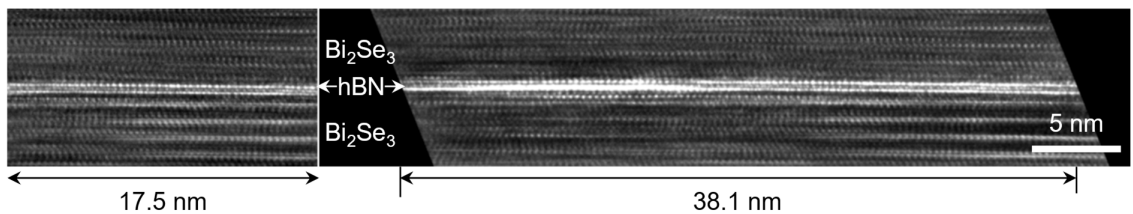

**Extended Data Fig. 2. Larger scale homogeneity of the Bi₂Se₃/hBN/Bi₂Se₃ interface.** Cross-sectional HR-TEM images of Bi$_2$Se$_3$/2ML hBN/Bi$_2$Se$_3$ taken along the zone axis [1$\bar{2}$10] exhibiting atomically sharp and clean interfaces.

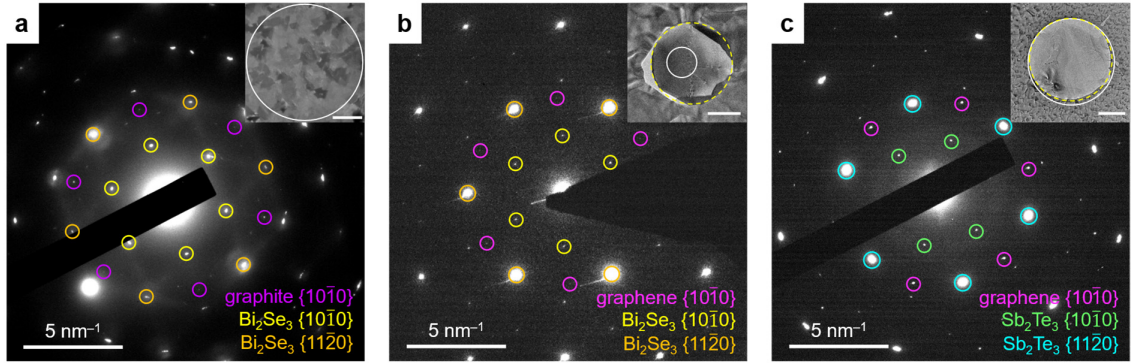

**Extended Data Fig. 3. Epitaxial relationship of the TI thin films grown on a single surface of suspended graphene substrates. a–c**, Plan-view SAED patterns of the $Bi_2Se_3$ thin film grown on the top surface of suspended graphite (**a**), the $Bi_2Se_3$ thin film grown on the bottom surface of suspended monolayer graphene (**b**), and the $Sb_2Te_3$ thin film grown on the top surface of suspended monolayer graphene (**c**). Insets: Plan-view TEM BF images. The white solid circles and the yellow dashed circles indicate the areas in which the SAED in the corresponding main panels are taken and the suspended regions, respectively. The entire area displayed in the inset is suspended for **a**. Scale bars in the insets: 1 μm (**a**), 200 nm (**b**), and 200 nm (**c**). All data (**a**–**c**) exhibit the preferential epitaxial relationship of $\{10\bar{1}0\}_{Bi_2Se_3 \text{ or } Sb_2Te_3} \parallel \{10\bar{1}0\}_{graphene \text{ or } graphite}$, indicating that the epitaxial alignment in the double-sided vdW epitaxy is owing to the local interaction at the interface with the suspended vdW substrate layer.

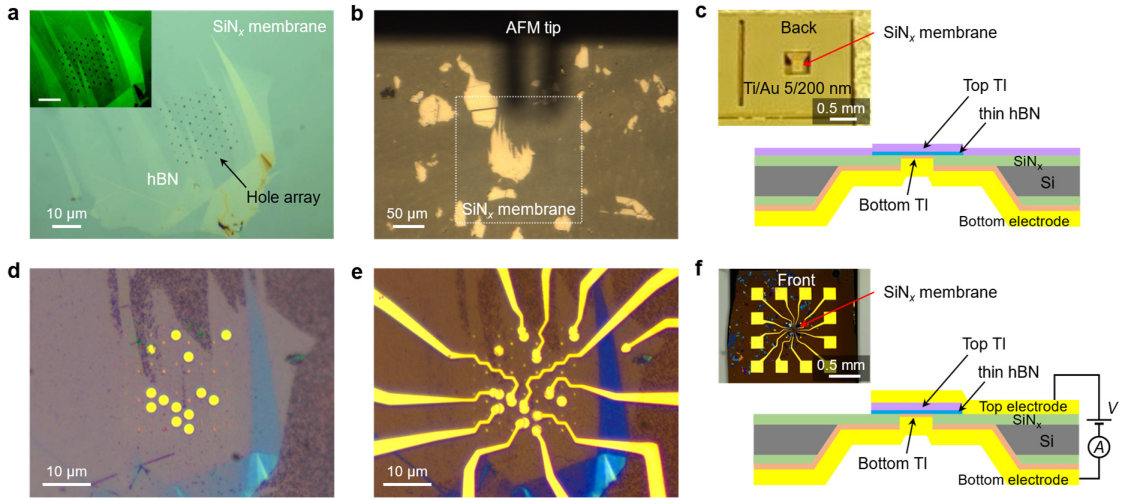

**Extended Data Fig. 4. Tunneling device fabrication processes. a**, Optical microscope image of atomically thin hBN layers suspended on the perforated SiN$_x$ membrane. Inset: High-contrast image of the main panel. Scale bar: 10 μm. **b**, Optical microscope image of the sample in (**a**) after the double-sided vdW epitaxy of TI thin films (10 QL Bi$_2$Se$_3$ on the back side and 20 QL Sb$_2$Te$_3$ on the top side). **c**, Photograph (upper left) and schematic (lower right) of the double-sided vdW-epitaxial heterojunction after the evaporation of the Ti/Au 5/200 nm common bottom contact. **d,e**, Optical microscope images of the junction arrays: after the fabrication of Ti/Au 5/55 nm individual top contacts followed by Ar plasma etching (**d**) and after the formation of bonding pads and leads (Ti/Pd/Au 5/20/125 nm). **f**, Optical microscope image (upper left) and schematic structure (lower right) of the finished device.

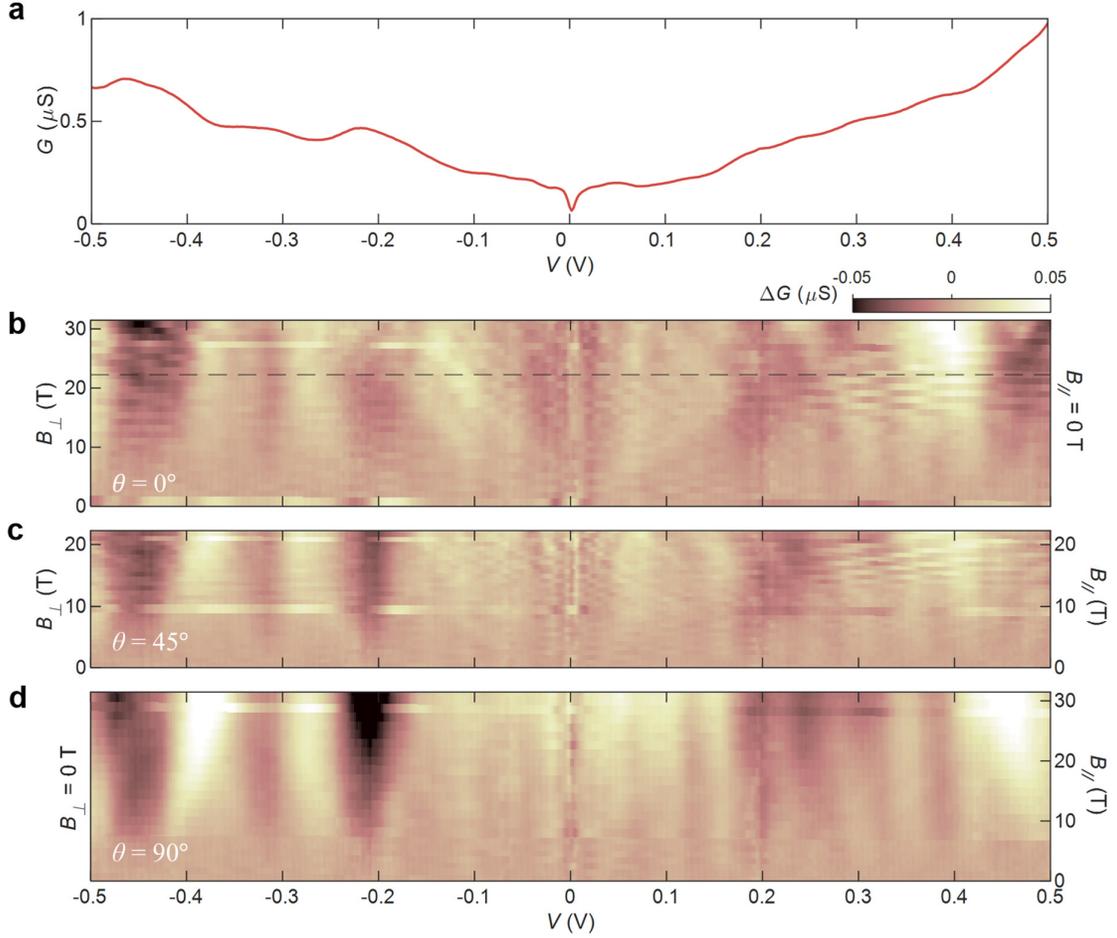

**Extended Data Fig. 5. Magnetic field-dependent tunnelling conductance of a Bi$_2$Se$_3$/3 ML hBN/Bi$_2$Se$_3$ device.**
**a**, Differential conductance $G(V) = \partial I/\partial V$ of the Bi$_2$Se$_3$/3ML hBN/Bi$_2$Se$_3$ device at zero magnetic field. **b–d**, Colour scale maps showing the field-modulated component of the conductance $\Delta G(V, B)$ at different field angles: out-of-plane (**b**), 45° tilted (**c**), and in-plane (**d**) directions. $B_\perp$ and $B_\parallel$ components of the field are indicated by the left and right ticks, respectively. The black dashed horizontal line in **b** indicates the maximum $B_\perp$ in **c**. We observe oscillatory features that develop at higher $B_\perp$, with their peak and valley positions varying as a function of $B_\perp$, in addition to the field-angle-independent oscillatory features arising from the bulk QWSs. We are however unable to conclusively identify conductance oscillations associated with the LL formation in TSSs, presumably because the conductance is dominated by the bulk-to-bulk tunnelling as both the b- and t-Bi$_2$Se$_3$ films are degenerately n-doped. The difference in thickness between the b-Bi$_2$Se$_3$ (10 QL) and t-Bi$_2$Se$_3$ (20 QL) is responsible for the asymmetry in the periodicity of the QWS-related oscillations at negative and positive biases. All measurements are performed at $T = 0.3$ K.

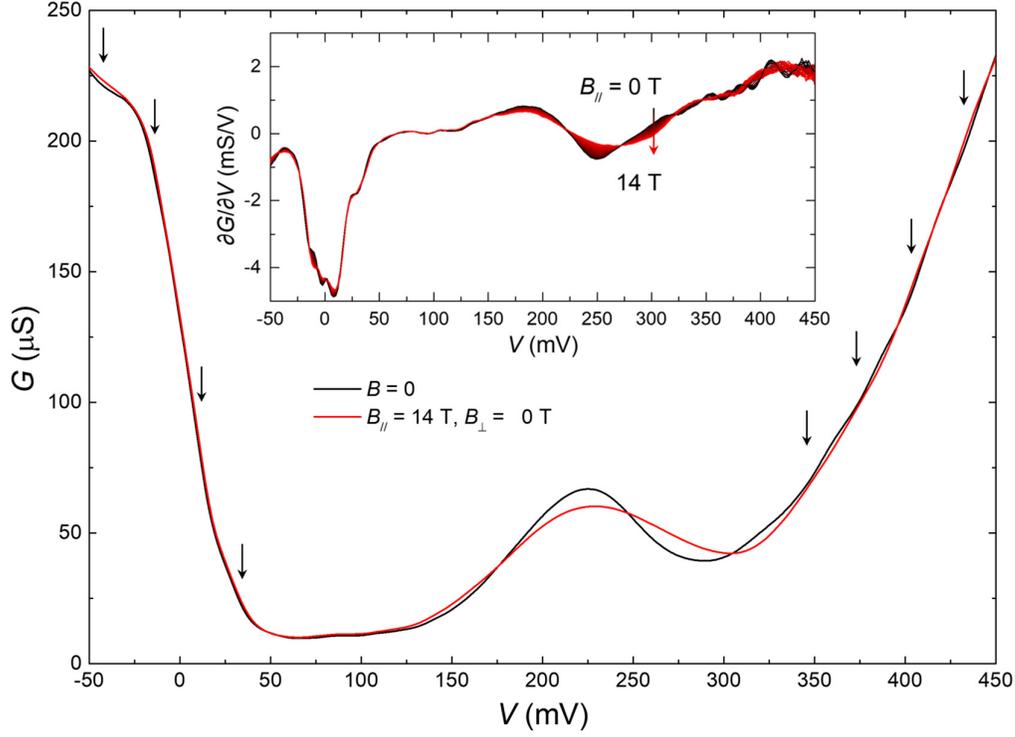

**Extended Data Fig. 6. Oscillations in the tunnelling conductance associated with the bulk QWSs.** Differential conductance $G(V) = \partial I/\partial V$ of the $Bi_2Se_3$/2ML hBN/$Sb_2Te_3$ device at zero magnetic field (black) and $B_\parallel$ = 14 T (red). The inset shows $\partial G/\partial V = \partial^2 I/\partial V^2$ as a function of $V$ at different $B_\parallel$, obtained by taking numerical derivative of $G(V)$. The black arrows indicate the positions of the peaks of the oscillatory feature (I) related to the bulk QWSs, appearing in $\Delta G(V, B)$ (Fig. 4b–e). One can find that they correspond to the valleys of the QWS resonances in $G(V, B = 0)$, which become smoothened out at higher $B_\parallel$ as fine structures develop due to the Zeeman effects. We also observe that the $G(V, B = 0)$ peak at $V$ = 225 mV, attributed to the alignment of bulk band edges, is suppressed at higher $B_\parallel$. All data is measured at $T$ = 1.8 K.

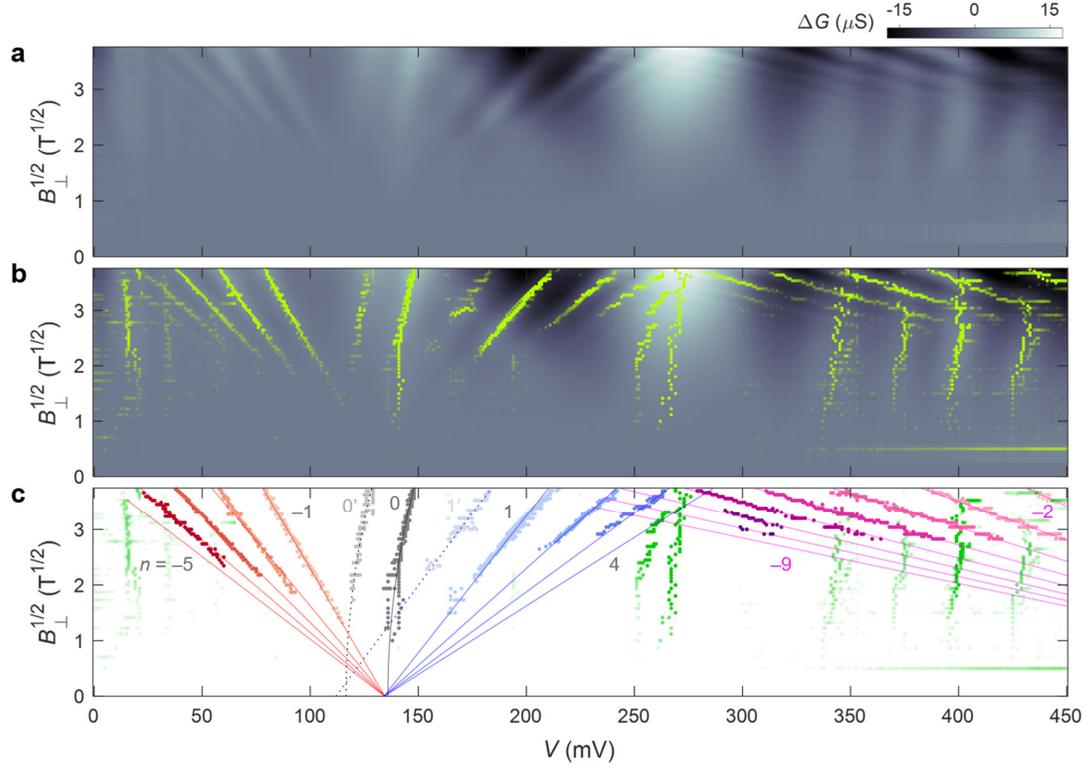

**Extended Data Fig. 7. LL peak identifications and LL index assignments. a**, Field-modulated component of the tunnelling conductance $\Delta G(V, B)$ at $\theta = 0°$ and $T = 1.8$ K, the same data as Fig. 4b replotted with $B_\perp^{1/2}$ as the vertical axis so that the LLs appear as linear lines. **b**, Peaks identified in **a**. The lime-coloured dots represent the peak positions identified by MATLAB's *findpeaks* function, a tool for locating local maxima, applied along the vertical and horizontal axes, as well as in two diagonal directions. The transparency of the dots corresponds to the prominence of the peaks the function returns. **c**, LL index assignments. The positions of the peaks assigned to each LL index $n$ are indicated by the corresponding colour. The green dots are the positions of the peaks not associated with the LLs. The solid lines tracing the peaks corresponding to $n < 0$ (red) and $n > 0$ (blue) LLs of $Sb_2Te_3$, and the inter-band inter-LL resonant tunneling between $Bi_2Se_3$ and $Sb_2Te_3$ (magenta) are reconstructed from the slopes and intercepts of the linear fits in Fig. 4f. The dark grey curve is a fit to the $n = 0$ LL peaks shifting linearly with respect to $B_\perp$. The dotted curve and line trace the $n = 0'$ and $1'$ LL peaks of unclear origin, respectively.